\documentclass[preprint,aps,showpacs]{revtex4}
\usepackage[dvips]{graphicx}

\begin{document}

\newcommand{\dfrac}[2]{\frac{\displaystyle #1}{\displaystyle #2}}
\preprint{VPI--IPPAP--02--08}

\title{Classical Implications of\\the Minimal Length Uncertainty Relation}
\author{
S\'andor~Benczik\footnote{electronic address: benczik@vt.edu},
Lay~Nam~Chang\footnote{electronic address: laynam@vt.edu},
Djordje~Minic\footnote{electronic address: dminic@vt.edu},
Naotoshi~Okamura\footnote{electronic address: nokamura@vt.edu}, 
Saifuddin~Rayyan\footnote{electronic address: srayyan@vt.edu}, and
Tatsu~Takeuchi\footnote{electronic address: takeuchi@vt.edu}
}
\affiliation{Institute for Particle Physics and Astrophysics,
Physics Department, Virginia Tech, Blacksburg, VA 24061}

\date{September 12, 2002}

\begin{abstract}

We study the phenomenological implications of the classical limit of
the ``stringy'' commutation relations
$[\hat{x}_i,\hat{p}_j]=i\hbar[(1+\beta\hat{p}^2)\delta_{ij}
+ \beta'\hat{p}_i\hat{p}_j]$.
In particular, we investigate the ``deformation'' of 
Kepler's third law and apply our result to
the rotation curves of gas and stars in spiral galaxies.
\end{abstract}

\pacs{02.40.Gh,11.25.Db,95.35.+d,98.35.Df}

\maketitle
\section{Introduction}

In this note, we continue our investigation \cite{Chang:2001kn,Benczik:2002tt}
of the phenomenological implications of 
``stringy'' commutation relations \cite{Kempf:1995su}
which embody the minimal length uncertainty relation
$\Delta x \ge (\hbar/2)( \Delta p^{-1} + \beta\,\Delta p)\;$
of perturbative string theory \cite{gross}.
In particular, we study the classical limit of the ``deformed'' 
commutation relations 
\begin{eqnarray}
\left[\hat{x}_i,\hat{p}_j\right] 
& = & i\hbar \left\{ ( 1 + \beta \hat{p}^2 )\,\delta_{ij}
                         + \beta' \hat{p}_i \hat{p}_j
             \right\} \;,\cr
\left[\hat{p}_i,\hat{p}_j\right] 
& = & 0 \;,\cr
\left[\hat{x}_i,\hat{x}_j\right]
& = & i\hbar\,\frac{(2\beta-\beta') + (2\beta+\beta')\beta \hat{p}^2}
                  {(1+\beta \hat{p}^2) }
      \left( \hat{p}_i \hat{x}_j - \hat{p}_j \hat{x}_i \right)\;,
\label{Stringy}
\end{eqnarray}
leading to the following ``deformed'' Poisson brackets,
\begin{eqnarray}
\{x_i,p_j\} & = & ( 1 + \beta p^2 )\,\delta_{ij} + \beta' p_i p_j\;,\cr
\{p_i,p_j\} & = & 0 \;,\cr
\{x_i,x_j\} & = & \frac{(2\beta-\beta') + (2\beta+\beta')\beta p^2}
                       { (1+\beta p^2) }
                  \left( p_i x_j - p_j x_i \right)\;.
\label{Eq:Poisson1}
\end{eqnarray}
The classical Poisson bracket is required to possess the same properties 
as the quantum mechanical commutator, namely, 
it must be anti-symmetric, bilinear, 
and satisfy the Leibniz rules and the Jacobi Identity.   
These requirements allow us to derive the general form of our 
Poisson bracket for any functions of the coordinates and momenta 
as \cite{Benczik:2002tt}
\begin{equation}
\{F,G\} = 
\left( \frac{\partial F}{\partial x_i}
       \frac{\partial G}{\partial p_j}
     - \frac{\partial F}{\partial p_i}
       \frac{\partial G}{\partial x_j}
\right) \{ x_i, p_j \}
+ \frac{\partial F}{\partial x_i}
  \frac{\partial G}{\partial x_j}
  \{ x_i, x_j \}\;,
\label{Eq:Poisson2}
\end{equation}
where repeated indices are summed.
Thus, the time evolutions of the coordinates and momenta 
in our ``deformed'' classical mechanics are governed by
\begin{eqnarray}
\dot{x}_i & = & \{x_i,H\} 
\;=\; \phantom{-}\{x_i,p_j\}\,\frac{\partial H}{\partial p_j} 
    + \{x_i,x_j\}\,\frac{\partial H}{\partial x_j} \;,\cr
\dot{p}_i & = & \{p_i,H\}
\;=\; -\{x_i,p_j\}\,\frac{\partial H}{\partial x_j} \;.
\label{Eq:Poisson3}
\end{eqnarray}

In Ref.~\cite{Benczik:2002tt}, we analyzed the motion of objects in 
central force potentials subject to these equations
and found that orbits in $r^2$ and $1/r$ potentials 
no longer close on themselves when $\beta$ and/or $\beta'$ 
are non-zero. 
This allowed us to place a stringent limit on the value of
the minimal length from the observed precession of the 
perihelion of Mercury,
\begin{equation}
\hbar\sqrt{\beta} < 2.3\times 10^{-68}\;\mathrm{m}\;,
\label{MercuryLimit}
\end{equation}
which was 33 orders of magnitude below the Planck length.

The natural question to ask next is whether there exist
other ``deformations'' of classical mechanics due to
$\beta$ and/or $\beta'$ that are either 1) observable
even for such a small value of $\hbar\sqrt{\beta}$, or
2) lead to an even more stringent limit due to their absence.
In the following, we will look at the ``deformation'' of
Kepler's third law.
We will find that while the deformation is not observable 
on the scale of the solar system, it may be observable at 
galactic scales.
In fact, such an effect may have already been seen in the
rotation curves of gas and stars in spiral galaxies.

\section{Aspects of the Deformed Classical Mechanics}

\noindent

Before we investigate the deformation of Kepler's third law, 
it is worthwhile to discuss some peculiarities of this 
novel classical dynamics.

First, it is obvious that the Poisson bracket
$\{x_i,x_j\}$ in Eq.~(\ref{Eq:Poisson1})
is not invariant under translations since it is proportional
to $(\vec{x}\times\vec{p})_{ij}$.
For the special case of $\beta'=2\beta$, the Poisson bracket $\{x_i,x_j\}$
is zero to order $\mathcal{O}(\beta)$ and we can recover
approximate translational invariance \cite{Kempf:1995su}.

Second, it is not obvious how to extend the fundamental Poisson brackets to
multi-particle systems, nor how to define appropriate 
`canonical transformations' that leave the fundamental
Poisson bracket invariant, even in the case of
one-particle systems .
For instance, consider the motion of 2 bodies of equal mass in 1D. 
Let us assume that the dynamics will be described by two sets of
`canonical' variables, $(x_1,p_1)$ and $(x_2,p_2)$, which satisfy
\begin{eqnarray}
\{x_i,p_j\} & = & \left(1 + \beta p_i^2\right)\delta_{ij}\;,\cr
\{p_1,p_2\} & = & \{x_1,x_2\} \;=\; 0\;.
\label{Eq:Poisson4}
\end{eqnarray}
Compare this expression to the multi-dimensional case, Eq.~(\ref{Eq:Poisson1}).
The right hand side of the first line is not $(1+\beta p^2)\,\delta_{ij}$,
where $p^2=\sum_i p_i^2$, but $(1+\beta p_i^2)\,\delta_{ij}$.  
This allows us to assume $\{x_1,x_2\}=0$.
If we change the variables naively to those of the center of mass frame,
\begin{equation}
\begin{array}{ll}
x_\mathrm{cm} = \dfrac{x_1+x_2}{2}\;, & p_\mathrm{cm} = p_1 + p_2\;,\\ 
x_\mathrm{rel} = x_1 - x_2\;,        & p_\mathrm{rel} = \dfrac{p_1-p_2}{2}\;,
\end{array}
\end{equation}
then we find that the sets $(x_\mathrm{cm},p_\mathrm{cm})$ and 
$(x_\mathrm{rel}, p_\mathrm{rel})$ are no longer `canonical'.
Of course, in this deformed dynamics momentum is no 
longer equal to \textit{mass}$\,\cdot\,$\textit{velocity} in general, 
so it is not surprising that the `canonical' momentum of the 
center of mass, if it exists, is not equal to the sum of momenta 
of the individual masses.

Third, consider 1D motion with the Hamiltonian given by
\begin{equation}
H = \frac{p^2}{2m} + V(x)\;.
\end{equation}
We will not consider any deformations of the Hamiltonian itself
though it is conceivable that some sort of modification may be
required for the consistency of the theory.
The equations of motion read
\begin{eqnarray}
\dot{x} & = & \{x,H\} \;=\; (1 + \beta p^2)\frac{p}{m}\;,\cr
\dot{p} & = & \{p,H\} \;=\; 
(1 + \beta p^2)\left(-\frac{\partial V}{\partial x}\right)\;.
\end{eqnarray}
As mentioned above, the momentum $p$ is no longer equal to $m\dot{x}$.  
From these equations, we can derive
\begin{equation}
(1+\beta p^2)(1+3\beta p^2) F = m\ddot{x}\;,
\label{ModFma}
\end{equation}
where
\begin{equation}
F\equiv -\frac{\partial V}{\partial x}\;.
\end{equation}
Thus we obtain a deformation of Newton's second law $F=m\ddot{x}$.
Now, notice that if the force $F$ is gravitational and proportional to
the mass $m$, the acceleration $\ddot{x}$ is not mass-independent
as usual because of the residual $m$-dependence through the momentum $p$.
Therefore, in Eq.~(\ref{ModFma}) the equivalence principle 
is dynamically violated.

It should be noted that the equivalence principle is already known 
to be violated in the context of perturbative string theory \cite{mende}. 
Fundamental strings, due to their extended nature, are subject to
tidal forces and do not follow geodesics.
However, in contrast to the current discussion where the
systems under consideration are of macroscopic dimensions, 
the violation discussed in Ref.~\cite{mende} was microscopic in nature.
Whether this is another example of UV/IR correspondence in string theory
remains to be seen.

\section{Motion in Central Force Potentials}

Let us now consider the motion of an object in a general central 
force potential where the Hamiltonian is given by
\begin{equation}
H = \frac{p^2}{2m} + V(r)\;,\qquad r = \sqrt{x_i x_i}\;.
\label{Hamil}
\end{equation}
We will apply the results of this section later to the case
when $V(r)$ is the gravitational potential and derive 
a ``deformed'' version of Kepler's third law.

As shown in Ref.~\cite{Benczik:2002tt}, 
the rotational symmetry of the Hamiltonian
Eq.~(\ref{Hamil}) leads to the conservation of the `deformed' angular momentum
\begin{equation}
L_{ij} = \frac{x_i p_j - x_j p_i}{(1 + \beta p^2)}\;,
\end{equation}
which in turn implies that the motion will be confined to a 2-dimensional
plane.  Expressing the position of the object in the plane in polar
coordinates $(r,\phi)$, the equations of motion can be cast into the form
\begin{eqnarray}
\dot{r} 
& = & \frac{1}{m}[\,1+(\beta+\beta') p^2\,]\,p_r \;,\cr
\dot{p}_r 
& = & \frac{1}{mr}[\,1 + ( \beta + \beta' ) p^2\,]
\left\{ ( p^2 - p_r^2 )
     - \left( mr\frac{\partial V}{\partial r} \right)
       \left( \frac{ 1 - \beta p^2 + 2\beta p_r^2 }{ 1 + \beta p^2 } 
       \right)
\right\}\;,\cr
\dot{\phi}
& = & \frac{L}{mr^2}
      \left\{ [\,1+(\beta+\beta')p^2 \,] (1+\beta p^2)
            + [\,(2\beta-\beta') + (2\beta+\beta')\beta p^2 \,]
              \left( mr\frac{\partial V}{\partial r} \right)
      \right\}\;,
\label{randphiEq}
\end{eqnarray}
where
\begin{eqnarray}
p_r & \equiv & \frac{(p\cdot x)}{r}
\;=\; \sqrt{ p^2 - \frac{L^2 (1+\beta p^2)^2}{r^2} }\;,\cr
L^2 & \equiv & -\frac{1}{2}L_{ij}L_{ji}
\;=\; \frac{ p^2\,r^2 - (p\cdot x)^2 }{ (1+\beta p^2)^2 }\;.
\label{prandL}
\end{eqnarray}
We restrict our attention to circular orbits and impose the constraints 
$\dot{r}=0$ and $\dot{p}_r=0$.  These conditions lead to
\begin{equation}
L = \frac{ p\,r }{(1+\beta p^2)}\;,
\label{cond3}
\end{equation}
and
\begin{equation}
\left( mr\frac{\partial V}{\partial r} \right)
= p^2 
\left( \frac{ 1 + \beta p^2 }{ 1 - \beta p^2 } \right)\;.
\label{p2eq}
\end{equation}
The radius of the orbit  $r$, and the magnitude of the momentum
$p=\sqrt{p^2}$ are now constants of motion.
So is the angular velocity $\omega \equiv \dot{\phi}$ which can
be expressed as
\begin{eqnarray}
\omega 
& = & \frac{p}{mr}
      \left\{ [\,1 + (\beta + \beta') p^2\,]
            + \frac{ (2\beta-\beta') + (2\beta+\beta') \beta p^2 }
                   { (1 + \beta p^2) }
              \left( mr\frac{\partial V}{\partial r} \right)
      \right\} \cr
& = & \frac{p}{mr}
      \left\{ [\,1 +(\beta + \beta')p^2\,]
            + \frac{ (2\beta-\beta') + (2\beta+\beta') \beta p^2 }
                   { (1 - \beta p^2) }\, p^2
      \right\} \cr
& = & \left(\frac{p}{mr}\right)
      \frac{(1+\beta p^2)^2}{(1-\beta p^2)}\;.
\label{omegaeq}
\end{eqnarray}
Solving Eq.~(\ref{p2eq}) for $p$ and substituting into
Eq.~(\ref{omegaeq}) yields
\begin{equation}
m^2 v^2 = m^2 r^2 \omega^2 =
\left( mr\frac{\partial V}{\partial r} \right) f(\varepsilon)\;,
\label{v2eq}
\end{equation}
where
\begin{eqnarray}
f(\varepsilon)
& \equiv &  \frac{1}{2}
                 \Biggl[\, ( 1 + 4\varepsilon - \varepsilon^2 )
                                  + ( 1 + \varepsilon )
                        \sqrt{ 1 + 6\varepsilon + \varepsilon^2 }
                 \,\Biggr] \;,\cr
\varepsilon 
& \equiv &
\beta mr\frac{\partial V}{\partial r} \;.
\end{eqnarray}
The function $f(\varepsilon)$ is a monotonically increasing 
function of $\varepsilon$ which behaves as
\begin{equation}
f(\varepsilon)
= \left\{ \begin{array}{ll}
               1 + 4\varepsilon  + \mathcal{O}(\varepsilon^2) & 
               \mbox{when $\varepsilon\ll 1$}\;, \\
               4\varepsilon + \mathcal{O}(\varepsilon^{-1})    & 
               \mbox{when $\varepsilon\gg 1$}\;.
              \end{array}
   \right.
\end{equation}
The graph of $f(\varepsilon)$ is shown in Fig.~1.

Note that $\beta'$ does not appear in our expression for $v^2$.
This means that the orbital period $T=2\pi r/v$ is not affected by $\beta'$. 
In particular, it is independent of whether approximate translational 
invariance ($\beta'=2\beta$) is realized or not.
Note also that $\varepsilon$, and thus the
deformation factor $f(\varepsilon)$, depends not only on $r$ but
also on $m$.  
As was the case with Eq.~(\ref{ModFma}),
this last fact can be interpreted as the dynamical violation
of the equivalence principle when our formalism is applied to
motion in gravitational fields.

\section{Kepler's Third Law}

For a spherically symmetric mass distribution, the gravitational
force on a mass $m$ at a distance $r$ from the center is given by
\begin{equation}
\frac{\partial V}{\partial r}
= m\,\frac{GM(r)}{r^2}\;,
\end{equation}
where $G$ is Newton's gravitational constant, and
$M(r)$ is the total mass enclosed within the radius $r$.
Thus, Eq.~(\ref{v2eq}) leads to the relation
\begin{equation}
v^2 = \frac{GM(r)}{r}\,f(\varepsilon)\;,
\label{Kepler}
\end{equation}
with
\begin{equation}
\varepsilon
= \beta m^2 \frac{GM(r)}{r}\;.
\end{equation}
In terms of the orbital period $T=2\pi r/v$,
we obtain
\begin{equation}
\frac{r^3}{T^2} = \frac{GM(r)}{4\pi^2}\,f(\varepsilon)\;.
\label{Kepler2}
\end{equation}
This is the `deformed' version of Kepler's third law.


\begin{table}
\begin{center}
\begin{tabular}{|c||c|c|c|c|c|c|} \hline
Planet 
& mass 
& eccentricity 
& semi-major axis 
& orbital period 
& apparent 
& predicted \\
& $m$ ($10^{24}$kg) 
& $e$ 
& $a$ ($10^{11}$m) 
& $T$ ($10^7$s) 
& $\left(\dfrac{a^3}{T^2}\dfrac{4\pi^2}{GM_\odot}\right)-1$
& $f(\varepsilon)-1$ \\
\hline\hline
Venus   
& $4.869$ 
& $0.00677323$ 
& $1.0820893$ 
& $1.9414149$  
& $+1\times 10^{-6}$ 
& $6\times 10^{-9}$ \\
\hline
Earth   
& $5.974$
& $0.01671022$ 
& $1.4959789$ 
& $3.1558149$ 
& $+3\times 10^{-6}$ 
& $6\times 10^{-9}$ \\
\hline
Mars    
& $0.6419$
& $0.09341233$ 
& $2.2793664$ 
& $5.9355036$
& $-6\times 10^{-5}\;$ 
& $5\times 10^{-11}$ \\
\hline
Jupiter 
& $1899$
& $0.04839266$ 
& $7.7841203$ 
& $37.435566$
& $+1\times 10^{-3}$ 
& $1\times 10^{-4}$ \\
\hline
Saturn  
& $568.5$
& $0.05415060$ 
& $14.267254$ 
& $92.929236$
& $+4\times 10^{-4}$ 
& $6\times 10^{-6}$ \\
\hline
Uranus  
& $86.8$
& $0.04716771$ 
& $28.709722$ 
& $265.13700$
& $+1\times 10^{-3}$ 
& $7\times 10^{-8}$ \\
\hline
Neptune 
& $102$
& $0.00858587$ 
& $44.982529$ 
& $520.04186$
& $+1\times 10^{-3}$ 
& $6\times 10^{-8}$ \\
\hline
\end{tabular}
\end{center}
\caption{The apparent and predicted violation of Kepler's third
law in planetary orbits.
}
\label{PLANETS}
\end{table}


Let us estimate how large this deformation can be for the 
planets in our solar system.
Using the data from Ref.~\cite{AstroData}, we calculate
the deformation factor $f(\varepsilon)$ for the planets
setting $\beta$ equal to the upper bound provided 
in Eq.~(\ref{MercuryLimit}).
The results are shown in Table~\ref{PLANETS}.

We also list the values of
\begin{equation}
\left( \frac{a^3}{T^2}\frac{4\pi^2}{GM_\odot} \right) - 1\;,
\end{equation}
which show the apparent violation of Kepler's third law.
Note that for elliptical orbits, the radius $r$ must be replaced
by the semi-major axis $a$.
(We have not included Mercury or Pluto in our list since the
eccentricities of these planets are rather large and the circular
orbit approximation is poor.)

As can be seen from Table~\ref{PLANETS}, the expected size of the
deviation due to our deformed dynamics is orders of magnitude below
the apparent deviation which can be explained by classical
Newtonian dynamics as a result of perturbations 
of the planets on each other.  
Even in the case of Jupiter, for which the predicted
deviation is the largest, the deviation due to the deformation is
an order of magnitude smaller than the apparent deviation.

Therefore, we can conclude that as long as the constraint
given in Eq.~(\ref{MercuryLimit}) is fulfilled,
any deformation of Kepler's third law due to non-zero 
$\beta$ is completely hidden under apparent violations due to more 
conventional effects.

\section{Rotation Curves of Spiral Galaxies}

The deformation of Kepler's third law will affect the 
$r$-dependence of $v^2$.
So let us consider whether our deformed dynamics
could provide a non-dark matter solution to the halo dark matter
problem.  

The halo dark matter problem pertains to the rotation curves 
of interstellar gas in spiral galaxies.
`Rotation curve' refers to the plot of $v^2$ as a function of
the distance $r$ from the galactic center.
If gravity due to the visible stars is the only force responsible 
for the centripetal acceleration, then the observed distribution of
stars and Kepler's third law tell us that $v^2$ should fall off 
asymptotically as $\sim 1/r$ as $r$ is increased.  
However, the actual observed rotation curves approach constant 
values instead, suggesting the existence of dark matter halos to 
account for the missing mass \cite{halo1}.

In our deformed dynamics, the correction term $f(\varepsilon)$
does increase the value of $v^2$ for a given mass distribution. 
Unfortunately, the deformation parameter 
$\varepsilon = \beta m^2 GM(r)/r$ itself falls off as
$\sim 1/r$ with $r$, so the correction vanishes for large $r$ and
cannot explain the plateau of the rotation curves.
What our formalism does predict, however, is that the size of
the correction will be different for different masses 
due to the $m$-dependence of $\varepsilon$: larger masses
will rotate at higher speeds at the same radius $r$,
in clear violation of the equivalence principle.

Curiously, it has been recently observed that the rotation curves
of stars and interstellar gas do not in general agree
when measured separately \cite{halo1,halo2}.
Could our `deformed' dynamics account for this difference?
For most galaxies it turns out that the gas rotates at 
higher speeds than the massive stars.
This may have a natural explanation by positing that the gas interacts 
differently with the dark matter and magnetic fields as
compared to the stars.
However, for two exceptional galaxies
reported in Ref.~\cite{halo2}, the stars are rotating at
much higher speeds than the gas.   
For the galaxy NGC7331, the speeds of the stars and gas
at distance $R_e/4$ from the center of the galaxy
($R_e$ is the radius which encloses half of the visible light from 
the galaxy) are reported to be
\begin{eqnarray}
v_\mathrm{gas}  & = & \phantom{0}79\pm 20\;\mathrm{km/s}\;, \cr
v_\mathrm{star} & = & 186\pm 10\;\mathrm{km/s}\;.
\end{eqnarray}
If we interpret this galaxy as a case where the 
mechanisms that would normally accelerate the gas over the 
stars are absent and the effect of `deformed' dynamics is manifest,
we can ask what value of $\beta$ would account for this difference. 
Since
\begin{eqnarray}
v^2 
& = & \frac{GM(r)}{r} 
      \left\{ 1 + 4\beta m^2 \frac{GM(r)}{r} + \cdots \right\} \cr
& = & \frac{GM(r)}{r}
      \left( 1 + 4\beta m^2 v^2 + \cdots \right)\;,
\end{eqnarray}
then
\begin{equation}
\frac{v_\mathrm{star}^2}{v_\mathrm{gas}^2}
\approx 1 + 4\beta m_\mathrm{star}^2 v_\mathrm{star}^2\;,
\end{equation}
where we have neglected the correction term for the gas.
If we assume the solar mass
\begin{equation}
M_\odot = 2\times 10^{30}\,\mathrm{kg}\;,
\end{equation}
as a typical mass of a star in this galaxy, we find
\begin{equation}
\hbar\sqrt{\beta} \sim 10^{-69}\,\mathrm{m}\;,
\end{equation}
which is well within the limit given by Eq.~(\ref{MercuryLimit})
and thus compatible with observations on the scale of the solar system.

\section{Conclusion and Discussion}

In this note, we have investigated possible observable consequences of 
the ``deformed'' Kepler's third law as implied by the classical limit 
of the ``stringy'' commutation relations, Eq.~(\ref{Stringy}),
which were in turn based on the minimal length uncertainty relation 
of perturbative string theory \cite{gross}.

Due to the stringent limit on $\beta$, Eq.~(\ref{MercuryLimit}), obtained 
from the precession of the perihelion of Mercury in a previous publication 
\cite{Benczik:2002tt}, the predicted sizes of the deviations for the 
planets are too small to be observed. 
However, the effect of the deformation can be
amplified at galactic scales due to the large momenta involved
and lead to significant changes in the rotation curves of stars
in spiral galaxies.

It would be interesting to contrast our deformed classical
mechanics with another modification of
Newton's theory, MOND (Modified Newtonian Dynamics)~\cite{MOND},
which apparently succeeds in predicting a flat rotation
curve by modifying Newton's second law to
\begin{equation}
F \sim m \,\frac{|\ddot{x}|}{a_0}\, \ddot{x}\;,
\end{equation}
at very small accelerations 
($\ddot{x} \ll a_0 \sim 10^{-10}\,\mathrm{m/s^2}$).
Our deformed dynamics predicts different
rotation curves for different masses, 
so it is apparently quite different from MOND which
does not violate the equivalence principle.
Nevertheless, it is interesting to ask whether there
exists a deformation of the usual Hamiltonian
dynamics which incorporates some kind of UV/IR relation and
is in the same universality class as MOND.
We hope to address this question in the future.


\acknowledgments

We would like to thank Vishnu Jejjala, Achim Kempf,
Yukinori Nagatani, Makoto Sakamoto, Hidenori Sonoda, and Joseph Slawny 
for helpful discussions, and Alan Chamberlin for his help with
the planetary data.
This research is supported in part by a grant from the U.S.
Department of Energy, DE--FG05--92ER40709, Task~A.

\bigskip
\bigskip

\begin{figure}[ht]
\begin{center}
\includegraphics{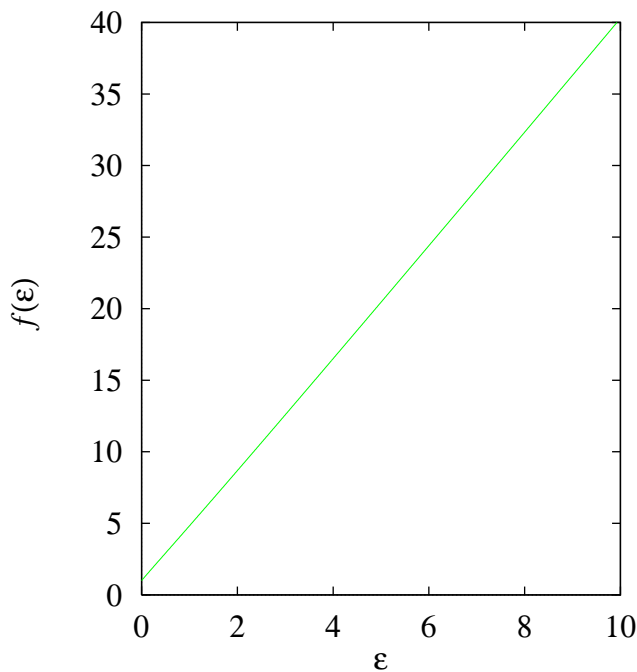}
\caption{The behavior of the function $f(\varepsilon)$.}
\label{fgraph}
\end{center}
\end{figure}



\end{document}